\documentclass[useAMS,usenatbib]{mn2e}
\usepackage[dvipdfmx]{graphicx}
\usepackage{amsmath,amssymb}
\usepackage{color}
\usepackage{natbib}

\begin{document}
\title[EM Counterpart to neutron star mergers]{Ultrarelativistic
electromagnetic counterpart to binary neutron star mergers}

\author[K. Kyutoku, K. Ioka and M. Shibata]{Koutarou Kyutoku$^1$,
Kunihito Ioka$^{2,3}$ and Masaru Shibata$^4$\\
$^1$Department of Physics, University of Wisconsin-Milwaukee, PO Box
413, Milwaukee, WI 53201, USA\\
$^2$Theory Center, Institute of Particles and Nuclear Studies, KEK,
Tsukuta 305-0801, Japan\\
$^3$Department of Particle and Nuclear Physics, the Graduate University
for Advanced Studies (Sokendai), Tsukuba 305-0801, Japan\\
$^4$Yukawa Institute for Theoretical Physics, Kyoto University, Kyoto
606-8502, Japan
}
\date{\today}

\maketitle

\begin{abstract}
 We propose a possibility of ultrarelativistic electromagnetic
 counterparts to gravitational waves from binary neutron star mergers at
 nearly all the viewing angles. Our proposed mechanism relies on the
 merger-shock propagation accelerating a smaller mass in the outer parts
 of the neutron star crust to a larger Lorentz factor $\Gamma$ with
 smaller energy $\sim 10^{47} \Gamma^{-1}$ erg. This mechanism is
 difficult to resolve by current 3D numerical simulations. The outflows
 emit synchrotron flares for seconds to days by shocking the ambient
 medium. Ultrarelativistic flares shine at an early time and in
 high-energy bands, potentially detectable by current X-ray to radio
 instruments, such as {\it Swift} XRT and Pan-STARRS, and even in low
 ambient density $\sim 10^{-2} \, \mathrm{cm}^{-3}$ by EVLA. The flares
 probe the merger position and time, and the merger types as black
 hole--neutron star outflows would be non-/mildly relativistic.
\end{abstract}


\section{Introduction} \label{sec:intro}

Binary neutron star (BNS) mergers are main sources of gravitational
waves (GWs) for ground-based laser-interferometric detectors, such as
advanced LIGO, advanced Virgo and KAGRA in the coming five years
\citep{ligo2010,lcgt2010,virgo2011}. GW detection will open a new window
for astronomy, and we will be able to test the theory of gravitation and
to probe the supranuclear-density matter in neutron stars
(NSs). Statistical studies suggest that a few tens of merger events are
observed in a year within a few hundred Mpc distance
\citep{ligovirgo2010}.

A simultaneous detection of electromagnetic (EM) signals is
indispensable for declaring a confident discovery of GWs
\citep{metzger_berger2012,piran_nr2013}. Since `hearing' a sound of GWs
entails a bad localization about degree$^2$ at best, `seeing' EM
counterparts will not only increase GW sensitivity but also expand {\it
multi messenger astronomy} by extracting information such as the host
galaxy and its redshift.

Short $\gamma$-ray bursts (SGRBs) are plausible counterparts to BNS
mergers \citep{nakar2007}. GWs will verify the merger hypothesis for
SGRBs. However, some observations suggest that SGRBs are beamed into a
small angle \citep{fong_etal2012}. Most SGRBs are off-axis and
undetectable to us, albeit GW observation is biased towards the binary's
rotational axis (i.e., probably the jet axis). The `orphan' afterglow,
which is produced by the off-axis jet decelerated to a mildly
relativistic velocity at a later time, is also dim.

Two promising models have been proposed for nearly isotropic EM
counterparts. One is the macronova or kilonova which shines on $\sim$
days after the merger in UV--optical bands via radioactive decay of {\it
r}-process elements
\citep{li_paczynski1998,kulkarni2005,metzger_mdqaktnpz2010}. The other
is radio synchrotron emission from the collisions between the ejecta and
the ambient medium, like $\gamma$-ray burst (GRB) afterglows, after
$\sim$ years from the merger \citep{nakar_piran2011}. Both of them are
based on non-/mildly relativistic outflows with $\sim 0.2$--$0.3 c$
(roughly an escape velocity of the NS) from a compact binary merger. The
outflows can be produced by neutrino-driven wind
\citep{dessart_obrl2009}, magnetically driven wind
\citep{shibata_ski2011,kiuchi_ks2012}, tidal ejection
\citep{roberts_klr2011} and shock-wave ejection \citep{goriely_bj2011}
(see also below). Recent fully general relativistic simulations show
that the ejecta mass is $\gtrsim 10^{-3} M_\odot$ for a wide range of
parameters even without neutrino or magnetic effects
\citep{hotokezaka_kkosst2013}.

In this Letter, we suggest a possibility of nearly omnidirectional
ultrarelativistic counterparts to BNS mergers for the first time to our
knowledge except for the GRB. We consider shock waves produced right
after the BNS collision
\citep{sekiguchi_kks2011,paschalidis_es2012}. Shock waves are launched
from the heated NS core to the NS crust non-relativistically at first,
and accelerate to a relativistic velocity down a steep density gradient
in the NS crust. After the shock breakout from the surface, the shocked
material expands into a nearly vacuum region, converting the
shock-heated internal energy into kinetic energy. The resulting Lorentz
factor $\Gamma$ of the ejecta is larger for outer and less massive
parts. Such a transrelativistic acceleration has been discussed in the
context of supernovae
\citep{sakurai1960,johnson_mckee1971,matzner_mckee1999,tan_mm2001,pan_sari2006}.

We estimate the relativistic ejecta mass to be $\sim 10^{-7} \Gamma^{-2}
\, M_\odot$ for $\Gamma \gg 1$, and calculate the synchrotron radiation
from relativistic blast waves decelerated by the ambient medium and
energized progressively by the inner catching-up ejecta. The
ultrarelativistic nature makes the flare bright at an early time
(seconds--days) and in high-energy bands (X-ray--radio bands) in
contrast to the non-/mildly relativistic cases. We find flares are
detectable by current X-ray, optical and radio instruments, such as {\it
Swift} XRT, Pan-STARRS and EVLA for our fiducial case.

The counterpart signals the merger time more precisely than non-/mildly
relativistic ones. The counterpart could also enable us to distinguish
the merger types, because black hole--NS mergers are not likely to be
accompanied by strong shocks for ultrarelativistic outflows.

Current 3D numerical simulations of BNS mergers have not sufficiently
resolved the NS crust. Although the results for non-/mildly relativistic
ejecta are solid and the existence of shock waves is implied by the
heatup of the colliding region
\citep{sekiguchi_kks2011,paschalidis_es2012}; currently, it is not
feasible to follow a tiny mass to a ultrarelativistic velocity because
of numerical viscosity, artificial atmosphere and limited computational
resources, even for Newtonian gravity. Therefore, it is worthwhile to
highlight the impacts of ultrarelativistic outflows for motivating the
future well-resolved calculations.

\section{Acceleration} \label{sec:accel}

\begin{figure}
 \begin{tabular}{c}
  \includegraphics[width=\columnwidth]{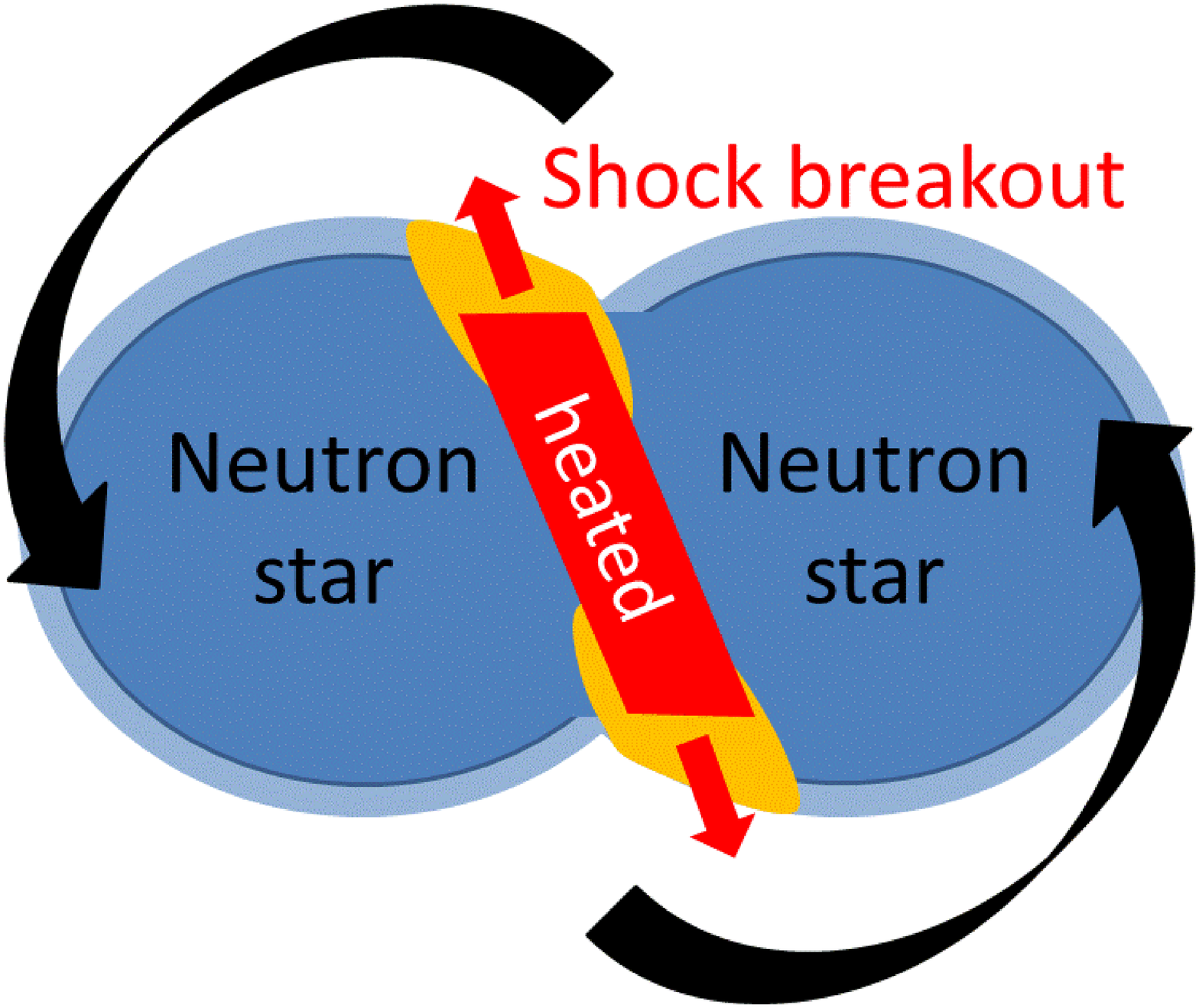} \\
  \includegraphics[width=\columnwidth]{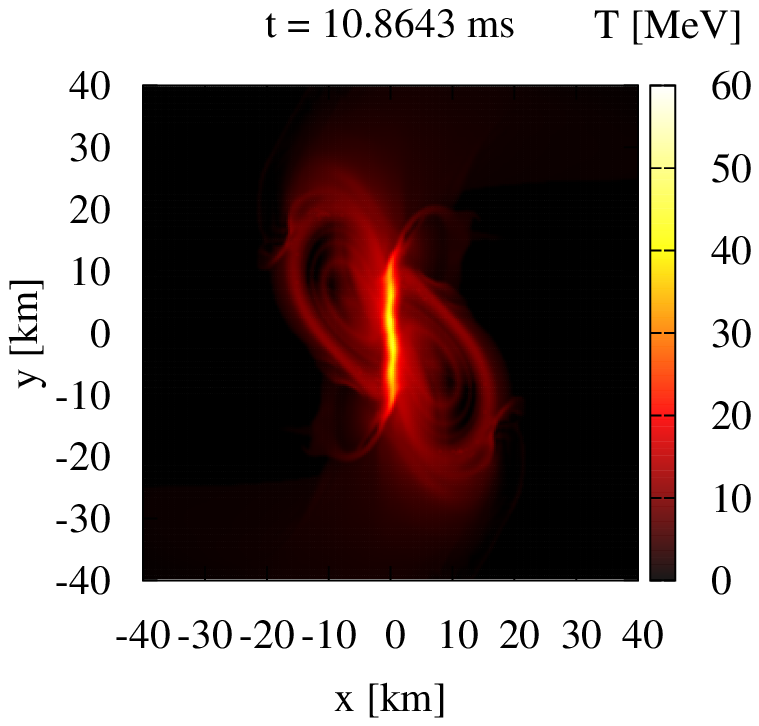}
 \end{tabular}
 \caption{Top: a schematic picture of the shock generation, propagation
 and mass ejection right after the BNS merger. Two blue ellipses are the
 BNS with a low temperature and the low-density crust is depicted with
 light blue. A red region at the contact surface is the shock-heated
 region. The black arrows denote the BNS motion just before the
 merger. Shock waves are generated from the contact of the BNS. The
 shocks become strong in the NS crust and eject a part of the NS crust
 ultrarelativistically. Bottom: a snapshot of merging BNS with 1.5
 $M_\odot$ taken from a simulation in \citet{sekiguchi_kks2011}. The
 temperature on the equatorial plane is shown, and the contact surface
 is heated up to $\sim 50$ MeV.} \label{fig:schematic}
\end{figure}

We first consider the mass ejection right after the NS collision. The
NSs collide with each other due to the gravitational radiation
reaction. The colliding part is shock heated up to a temperature of
$\sim 50$ MeV. Because of an oblique collision, the shocked region has a
pancake-like shape with the thickness $R_\mathrm{sh} \sim O(1)$ km
\citep{sekiguchi_kks2011,paschalidis_es2012}, as shown in
Fig.~\ref{fig:schematic}.

The hot material in the colliding region expands towards a cold,
low-pressure region, i.e., from the heated NS core to the NS crust. The
striking difference of the pressure between them drives shock waves
propagating the NS crust towards an NS surface. The initial shock
velocity $v_\mathrm{ini}$ should be comparable to the sound velocity of
the core material $\sim 0.25 c$ \citep{oertel_fn2012}, where $c$ is the
speed of light. At this stage, the shocked material cannot escape from
the merged remnant, because the expanding velocity is less than the
escape velocity,
\begin{equation}
 v_\mathrm{esc} \approx 0.74 c \left( \frac{M_*}{2.8 M_\odot}
			       \right)^{1/2} \left( \frac{R_*}{15 \,
			       \mathrm{km}} \right)^{-1/2} ,
\end{equation}
where $M_*$ and $R_*$ are the mass and radius of the merged remnant,
respectively.

The shock is accelerated descending a steep density gradient in the NS
crust with the thickness $R_c \approx 1$ km. The density profile of the
crust is approximately given by $\rho \propto x^n$, where $\rho$ is the
rest-mass density, $x$ is the depth from the surface and $n$ is the
polytropic index of the crust equation of state (EOS). We adopt $n=3$ as
a fiducial value, because this is for the relativistic degenerate
electron gas and is consistent with more detailed nuclear-theory-based
EOSs \citep{chamel_haensel2008}.

The shock velocity increases as $\propto \rho^{-\alpha}$ with $\alpha
\approx 0.187$ for $n=3$ in the non-relativistic regime
\citep{sakurai1960}. Once the shock is accelerated beyond $\approx 0.5
v_\mathrm{esc}$, the shocked material can escape from the BNS by
converting thermal energy into kinetic energy to obtain $\approx
v_\mathrm{esc}$ later
\citep{sakurai1960,matzner_mckee1999}. Specifically, the shock velocity
increases by a factor of $0.5 v_\mathrm{esc} / v_\mathrm{ini} \sim 1.5$
when the density drops by $1.5^{- 1 / \alpha} \approx 0.1$. The crust
material outside this density can escape as ejecta. For $\rho \propto
x^n$, the ejecta mass is estimated to be
\begin{align}
 M_\mathrm{sh} & \approx M_\mathrm{c} \left( \frac{R_\mathrm{sh}}{R_*}
 \right) \left( \frac{0.5 v_\mathrm{esc}}{v_\mathrm{ini}}
 \right)^{-(n+1)/n \alpha} \nonumber\\
 & \approx 4.4 \times 10^{-5} M_\odot \left(\frac{v_\mathrm{ini}}{0.25
 c} \right)^{7.1}
 \left(\frac{v_\mathrm{esc}}{0.74 c} \right)^{-7.1},
\label{eq:Msh}
\end{align}
where $R_\mathrm{sh} / R_* \approx 1 \, \mathrm{km} / 15 \, \mathrm{km}$
is a geometrical fraction of the crust mass $M_\mathrm{c} \approx 0.01
M_\odot$ that is swept by the shock (Fig.~\ref{fig:schematic}).

The ejecta is approximately spherical. The reason for this is that the
shock is initially non-relativistic, and therefore expands into an angle
given by the inverse of the Lorentz factor, $O(1)$. No confinement
mechanism works. Since the ejecta geometry is not jet-like but annular,
where the annulus is ejected in the {\it yz} plane for the bottom panel
of Fig.~\ref{fig:schematic}, the solid angle is $2 \pi \times O(1)$.

The outer and less massive ejecta accelerates to a ultrarelativistic
velocity. Specifically, the shock attains a Lorentz factor $\Gamma_s
\sim 10$ when the density drops by a factor of $\sim \Gamma_s^{2 + 4 /
n\sqrt{3}} \sim 10^4$ within a thin layer of $\sim R_c / \Gamma_s^{( 2 +
4 / \sqrt{3} ) /n} \sim R_c / 30$
\citep{johnson_mckee1971,pan_sari2006}. Then, after the breakout from
the surface, the shocked material accelerates to a Lorentz factor
$\Gamma \sim \Gamma_s^{1+\sqrt{3}} \gtrsim 500$ by converting the
shock-heated internal energy into the kinetic energy with the aid of the
pressure of the inner ejecta. To resolve the thin layer of $\lesssim R_c
/ 30$ in mesh-based simulations, a grid size of $\lesssim 10$ m is
required. Such a high-resolution simulation is not feasible at present
and in the near-future.

To make a detailed estimate, we apply a transrelativistic acceleration
model of a supernova exploding the stellar envelope
\citep{tan_mm2001}. The kinetic energy of ejecta with a velocity above
$\beta \Gamma$, where $\beta c$ is the ejecta velocity, is given by
equation 56 of \citet{tan_mm2001} as \footnote{We adopt $f_\mathrm{sph}
= 0.85$ and $C_\mathrm{nr} = 2.03$ in \citet{tan_mm2001}.}
\begin{equation}
 E ( > \! \beta \Gamma ) = \left( \frac{v_\mathrm{esc}}{0.58 c}
			   \right)^{(n+1)/n \alpha} F ( \beta \Gamma )
 M_\mathrm{sh} c^2 . \label{eq:kinetic}
\end{equation}
An exact form of the distribution $F ( \beta \Gamma )$ is given by
equation 38 of \citet{tan_mm2001}. Neither gravity nor rotation is
expected to affect the shock and post-shock acceleration, since the
shock crosses the crust in $O(10) \mu \mathrm{s}$, which is much shorter
than the dynamical time-scale and rotational period at the mass-shedding
limit, $O(1)$ ms. The heatup of the stellar interior will take $O(100)
\mu \mathrm{s}$, which might require detailed modelling.

\begin{figure}
 \includegraphics[width=\columnwidth]{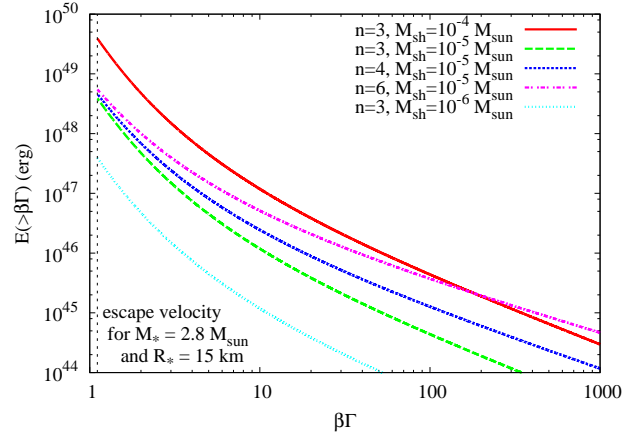} \caption{The kinetic
 energy distribution of ejecta with a velocity larger than $\beta
 \Gamma$ for various polytropic indices $n$ and the ejected mass by the
 shock breakout $M_\mathrm{sh}$. We assume $M_* = 2.8 M_\odot$ and $R_*
 = 15$ km.} \label{fig:kinetic}
\end{figure}

Fig.~\ref{fig:kinetic} shows the kinetic energy distribution of the
ejecta $E ( > \! \beta \Gamma)$. The energy of ultrarelativistic ejecta
$E ( > \beta \Gamma \approx \Gamma = 10)$ is $\gtrsim 10^{46}$ erg for
our fiducial case. For $\beta \Gamma \gg 1$, Eq.~\eqref{eq:kinetic}
yields
\begin{equation}
 E ( > \! \Gamma ) \approx 2.6 \times 10^{47} \; \mathrm{erg} \; (
  \Gamma^{-0.94} + \Gamma^{-0.20} )^{5.5} \left(
					   \frac{M_\mathrm{sh}}{4.4
					   \times 10^{-5} M_\odot}
					  \right) , \label{eq:fkinetic}
\end{equation}
where we assume $M_* = 2.8 M_\odot$, $R_* = 15$ km and $n=3$. The
high-$\Gamma$ component carries small but still appreciable energy for
the emission as $E ( > \! \Gamma ) \propto \Gamma^{-0.58-1.58/n} \sim
\Gamma^{-1.1}$, while the mass is tiny as $\sim 10^{-7}\Gamma^{-2.1} \,
M_\odot$. The energy distribution becomes harder for a larger value of
$n$, providing a possible way to infer the EOS of the NS crust in
principle.

The energy distribution is sensitive to the value of $M_\mathrm{sh}$,
and therefore $v_\mathrm{ini}$ and the polytropic index, $n$. The
density profile could be affected by the neutrino/magnetic wind, and the
shock acceleration will not work efficiently when the density has a
stellar-wind-like profile and does not go to zero rapidly. The breakdown
of plane-parallel approximation in \citet{tan_mm2001} could modify the
$\Gamma$ distribution (but see their section 2.5 for aspherical
explosions). Some part of the ultrarelativistic ejecta may be
decelerated before emission by surrounding material such as a
tidally elongated NS, especially for an unequal-mass binary. The
neutrino losses may decrease the acceleration pressure (but other
radiation components persist). The accurate estimation of
$M_\mathrm{sh}$ and the $\Gamma$ distribution taking these caveats is
left for future study.

The ejecta may be also accelerated when density waves propagate across
the entire core to the opposite surface. The amount of shock-breakout
ejection would be larger by an order of magnitude due to a larger
geometrical fraction. Whether density waves propagate across the core
depends on the NS structure, and thus on the EOS of the NS core.

\section{Radiation} \label{sec:rad}

Next, we calculate the spectra and light curves of EM signals applying
the synchrotron shock model of the GRB afterglow
\citep{sari_pn1998,ioka_meszaros2005}. The outflow generates a forward
shock sweeping the ambient medium with a constant number density
$n_\mathrm{H}$. A fraction $\epsilon_B \sim 0.01$ of the internal energy
released by the shock amplifies the magnetic field $B$, while a fraction
$\epsilon_\mathrm{e} \sim 0.1$ accelerates electrons with a Lorentz
factor distribution $\mathrm{d} N_\mathrm{e} / \mathrm{d}
\gamma_\mathrm{e} \propto \gamma_\mathrm{e}^{-p}$, where $N_\mathrm{e}$
and $\gamma_\mathrm{e}$ are the number and Lorentz factor of electrons,
respectively, for $\gamma_ \mathrm{e} \ge \gamma_m$ (a minimum value)
and $p \approx 2.2$.

Hereafter, we assume that the ejecta outspreads completely spherically,
and give the lower limit of luminosity. If the ejecta is concentrated
within an angle $\theta$ from the annular heated region, an observable
angle decreases by $\approx \theta / \pi$ but the isotropic energy
increases by $\approx \pi / \theta$ for a BNS merger, and finally
detection rates will increase by $\approx \sqrt{ \pi / \theta}$. Our
fiducial model can be detected up to $\gtrsim 200$ Mpc for optimal
observational bands and the value of $n_\mathrm{H}$, as shown later.

The outflow carries larger energy in inner, lower $\Gamma$ part in
Eqs.~\eqref{eq:kinetic} and \eqref{eq:fkinetic}. The slow flow rear-ends
and refreshes the external shock which is decelerated by the ambient
density \citep{rees_meszaros1998,sari_meszaros2000}. We illustrate
formulae for a power-law distribution of the kinetic energy,
\begin{equation}
 E ( > \! \Gamma ) = \tilde{E} \Gamma^{1-s} , \label{eq:single}
\end{equation}
where $\tilde{E} = 2.6 \times 10^{47}$ erg and $s = 2.1$ for our
fiducial case with $\Gamma \gg 1$, and $\tilde{E}$ and $s$ rise as
$\Gamma$ drops in equation \eqref{eq:fkinetic}. We also adopt an
ultrarelativistic approximation, $\Gamma \gg 1$. Note that $\gamma_m$
and $B$ are both proportional to $\Gamma$ in this approximation. Once
the fastest flow begins to decelerate, the catching-up condition for a
slower flow is given by $\tilde{E} \Gamma^{1-s} = 16 \pi \Gamma^2 R^3
n_\mathrm{H} m_\mathrm{p} c^2 / 17$
\citep{blandford_mckee1976,sari_meszaros2000}, where $m_\mathrm{p}$ is
the proton mass and $R$ is the shock radius. Since the shock radius is
connected with the observer time $t$ by $R \approx 4 \Gamma^2 c t$
\citep{sari_pn1998}, we obtain the hydrodynamic evolutions
\begin{equation}
 \Gamma (t) = ( c t / \ell_\mathrm{S} )^{- 3 / (s+7)} \; , \; R (t) = 4
  \Gamma (t)^2 c t ,
\end{equation}
where $\ell_\mathrm{S} \equiv ( 17 \tilde{E} / 1024 \pi n_\mathrm{H}
m_\mathrm{p} c^2)^{1/3}$ is the Sedov length. Note that $\Gamma \propto
t^{-0.33}$ and $R \propto t^{0.34}$ for our fiducial value $s=2.1$,
compared to $\Gamma \propto t^{-3/8}$ and $R \propto t^{1/4}$ for a
single-velocity shell.

Given the hydrodynamics above, we can calculate the evolution of
radiation \citep{sari_pn1998,ioka_meszaros2005}. The synchrotron flux
has a broken power-law spectrum, $F_\nu \propto \nu^{1/3}$ for $\nu <
\nu_m$ and $F_\nu \propto \nu^{- (p-1) / 2} \approx \nu^{-0.6}$ for $\nu
> \nu_m$, where $\nu_m \propto \gamma_m^2 \Gamma B \propto \Gamma^4
\propto t^{- 12 / (s+7)}$ is the characteristic synchrotron
frequency. The cooling frequency is high ($>$MeV) for typical
parameters. The self-absorption frequency is only relevant for the radio
band with $n_\mathrm{H} \gtrsim 1 \, \mathrm{cm}^{-3}$. For a given
frequency $\nu$, the flux reaches the maximum value $F_{\nu,m} \propto
N_\mathrm{e} \gamma_m^2 \Gamma^2 B^2 / \nu_m$ as $\nu_m$ crosses $\nu$,
with $N_\mathrm{e} \propto R^3$. The peak time and flux are given by
\begin{align}
 t_\mathrm{peak} & = 6.2 \; \mathrm{ms} \;
 \epsilon_{\mathrm{e},-1}^{1.52} \epsilon_{B,-2}^{0.38}
 \tilde{E}_{47}^{1/3} n_{\mathrm{H},0}^{0.05} \nu_{18}^{-0.76} , \\
  F_\mathrm{peak} & = 0.21 \; \mu \mathrm{Jy} \;
 \epsilon_{\mathrm{e},-1}^{0.55}\epsilon_{B,-2}^{0.64} \tilde{E}_{47}
 n_{\mathrm{H},0}^{0.64} D_2^{-2} \nu_{18}^{-0.28} ,
\end{align}
where $Q_x \equiv Q / 10^x$ in units of erg for $\tilde{E}$, Mpc for
$D$, cm$^{-3}$ for $n_\mathrm{H}$ and Hz for $\nu$. The flux grows as
$F_\nu \propto t^{(3s+1)/(s+7)} \approx t^{0.80}$ and decays as $F_\nu
\propto t^{3(s+1-2p)/(s+7)} \approx t^{-0.43}$ across the peak. These
light-curve behaviours could constrain the $\Gamma$ distribution, i.e.,
$s$, and hence the crust EOS, $n \approx 1.58 / ( s - 1.58 )$, in
principle. The degeneracy between $n$ and $p$ is solved by the
high-energy spectral index, $F_\nu \propto \nu^{-(p-1)/2}$.

\begin{figure}
 \includegraphics[width=\columnwidth]{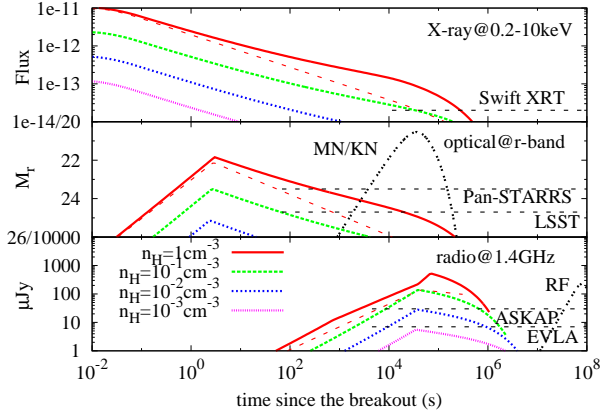} \caption{Light curves of
 the ultrarelativistic EM counterpart with $\tilde{E} = 2.6 \times
 10^{47}$ erg at 100 Mpc distance in the X-ray ($0.2$--$10$ keV
 integrated flux in $\mathrm{erg \, cm^{-2} \, s^{-1}}$), optical (in
 629 nm, {\it r}-band magnitude) and radio (1.4GHz in $\mu \mathrm{Jy}$)
 bands for various values of $n_\mathrm{H}$ down to $\Gamma \approx
 1$. Long dashed red curves are the analytic approximations with the
 power-law distribution in Eq.~\eqref{eq:single} with $n_\mathrm{H} = 1
 \, \mathrm{cm}^{-3}$. Short dashed black lines are sensitivity curves
 of {\it Swift} XRT, Pan-STARRS, LSST, ASKAP and EVLA. Double dotted
 black curves in the middle (labelled as MN/KN) and bottom (RF) panels
 show the non-relativistic EM counterparts of the macronova/kilonova
 \citep{li_paczynski1998} and the radio flare \citep{piran_nr2013},
 respectively, with the ejecta mass $10^{-3} M_\odot$, ejecta velocity
 $0.2 c$ and $n_\mathrm{H} = 1 \, \mathrm{cm}^{-3}$. For the
 macronova/kilonova, the heating efficiency and opacity are taken to be
 $3 \times 10^{-6}$ and $0.1~\mathrm{cm^2~g^{-1}}$, respectively
 \citep{metzger_mdqaktnpz2010}.} \label{fig:light}
\end{figure}

Fig.~\ref{fig:light} shows the light curves in X-ray, optical and radio
bands using the (non-power-law) velocity distribution in
Eqs.~\eqref{eq:kinetic} and \eqref{eq:fkinetic} with various ambient
densities $n_\mathrm{H}$. In contrast to the non-/mildly relativistic
cases, i.e., optical macronovae/kilonovae and radio flares, the
ultrarelativistic signals appear in the early phase down to seconds and
in the high frequency up to X-ray. The X-ray and optical peaks
correspond to $\Gamma \approx 400$ and 70, respectively.

\section{Discussion} \label{sec:dis}

The ultrarelativistic counterpart proposed here is bright from the
early epoch soon after the BNS merger, and decays rapidly. While this
feature is advantageous to confirm a tight association with GWs, the
observation will require efficient strategies. One such strategy is
full-time EM monitoring of nearby (up to $\sim 100$ Mpc) galaxies, where
the EM signals trigger the GW analyses like the SGRB case. The method is
expensive, but enables us to discover many other transients including
supernovae as a by-product.

The other strategy is prompt follow-up by EM instruments in response to
rapid alerts from GW detector networks. The localization requires at
least three and hopefully more than four GW detectors. Because a
localization error will be $\sim 1 \, \mathrm{degree}^2$ at best for a
BNS merger \citep{fairhurst2011}, covering this large area is crucial
for the EM follow-up. This will be challenging but not impossible. {\it
Swift} XRT has $0.15 \, \mathrm{degree}^2$ field of view (FOV). Tiling
the FOV will allow us to detect the decaying phase of the X-ray signal,
although the required number of tiles is $\gtrsim 10$. The latency from
GW detection to follow-up observation, which could be $\sim 12$ h
\citep{ligovirgoswift2012}, should be reduced as possible for efficient
tiling. Detecting the X-ray peak may be possible if EM precursors are
observed in advance (e.g., \citealp{ioka_taniguchi2000}). The optical
flare can be observed around its peak by all-sky surveys, such as
Pan-STARRS with $7 \, \mathrm{degree}^2$ FOV and LSST with $9.6 \,
\mathrm{degree}^2$ FOV, if $n_\mathrm{H} \gtrsim 10^{-1} \,
\mathrm{cm}^{-3}$. The radio flare can be also detected around the peak
by EVLA even in low ambient density $n_\mathrm{H} \approx 10^{-2} \,
\mathrm{cm}^{-3}$. The low ambient density is suggested by the radio
observations of SGRBs \citep{berger2010}. Since EVLA has relatively
small FOV of $0.25 \, \mathrm{degree}^2$, large FOV instruments such as
ASKAP with $30 \, \mathrm{degree}^2$ FOV may be more realistic
choices. Follow-up observation in optical and radio bands has also to be
performed as rapid as possible after GW detection to cover the
localization error efficiently during bright emission. Detecting the
short-lived emission proposed in this Letter will be more challenging
for typical localization errors, 10--$100 \; \mathrm{degree}^2$ dependent
on the detector network configuration, than for optimistic localization
errors, $\sim 1 \; \mathrm{degree}^2$.

We expect that these detectors will always find the emission by fully
covering the GW localization error region if it is $\sim 1 \;
\mathrm{degree}^2$, which might be possible for the merger at 100
Mpc. The detection probability may be estimated by the fraction of the
error region that follow-up observations can cover before the emission
fades away. Assuming $n_\mathrm{H} = 1 \; \mathrm{cm}^{-3}$, XRT, LSST
and EVLA will detect the emission up to $\sim 10^5, 10^4$ and $10^6$ s
after the merger at 100 Mpc, respectively. The number of available
pointings and the total FOV is estimated by comparing these values to
required integration time of each detector, and the probability is
found. By contrast, if the localization error is $\gtrsim 10 \;
\mathrm{degree}^2$ or the emission is dimmer due to a larger distance or
lower ambient density, XRT might find the emission only $\lesssim 10\%$
of the events. LSST and EVLA will be able to detect the emission even in
such cases. We would not like to be conclusive at this point, however,
due to enormous uncertainties associated with the ambient density, GW
localization errors including shapes of them, and the delay from GW
detection to follow-up observations.

We also speculate that GeV--TeV $\gamma$-rays could be generated via
inverse Compton scatterings or hadronic processes, such as {\it
p}--$\gamma$ collisions. If energy $E_\gamma$ is converted to
$\gamma$-rays with typical energy $e_\gamma$, an expected number of
photons $N_\gamma$ for a detector with area $A$ on the earth will be
\begin{equation}
 N_\gamma \approx 50 \left( \frac{E_\gamma}{10^{45} \, \mathrm{erg}}
		     \right) \left( \frac{100 \, \mathrm{GeV}}{e_\gamma}
			     \right) \left( \frac{A}{1 \, \mathrm{km}^2}
				     \right) D_2^{-2} .
\end{equation}
This suggests that km$^2$ future instruments such as CTA could also
detect EM signals in $\gamma$-rays. The TeV $\gamma$-rays are not
attenuated by the infrared background at $\sim 100$ Mpc.

Ultrarelativistic outflows could also be produced by other mechanisms
such as the Poynting wind from the NS surface with a small baryon load
like in the magnetar models for GRBs \citep{metzger_gtbq2011}. In this
case, the EM signals could arise from the magnetic reconnection without
the radioactivity or the ambient medium.

Before closing this section, we again summarize the caveats of our model
and necessary studies in the future. Our proposed ejection mechanism is
based on analytically idealized shock and post-shock acceleration. Local
plane-parallel geometry is assumed for a restricted region depicted in
Fig.~\ref{fig:schematic} to apply a model in \citet{tan_mm2001}, and the
ejecta is assumed to be isotropic relying only on the fact that the
shock is initially non-relativistic. The validity of these assumptions
has to be confirmed by numerical simulations with grid resolutions
$\lesssim 10$ m in the future. We neglected the gravity and rotation
based on the time-scale comparison, and this assumption requires more
quantitative validation by the simulations. Possible modification of the
crustal density profile and screening of ultrarelativistic ejecta by
other ejection mechanisms has also to be investigated.

\section*{Acknowledgements}

We are grateful to Akira Mizuta and Hajime Takami for valuable
discussion, and to Kenta Kiuchi for providing the bottom panel of
Fig.~\ref{fig:schematic}. We also thank our referee for helpful
comments. This work was supported by the Grant-in-Aid for Scientific
Research No.~21684014, 22244030, 24000004, 24103006, and 24244028 of
Japanese MEXT and JSPS Postdoctoral Fellowship for Research Abroad.

\bibliographystyle{mn2e}

\end{document}